# A Quartic Kernel for Pathwidth-One Vertex Deletion[*]

Geevarghese Philip[1], Venkatesh Raman[1], and Yngve Villanger[2]

[1]The Institute of Mathematical Sciences, Chennai, India.
{gphilip|vraman}@imsc.res.in

[2]University of Bergen, N-5020 Bergen, Norway.
yngve.villanger@uib.no

**Abstract**

The pathwidth of a graph is a measure of how path-like the graph is. Given a graph $G$ and an integer $k$, the problem of finding whether there exist at most $k$ vertices in $G$ whose deletion results in a graph of pathwidth at most one is NP-complete. We initiate the study of the parameterized complexity of this problem, parameterized by $k$. We show that the problem has a quartic vertex-kernel: We show that, given an input instance $(G = (V, E), k); |V| = n$, we can construct, in polynomial time, an instance $(G', k')$ such that (i) $(G, k)$ is a YES instance if and only if $(G', k')$ is a YES instance, (ii) $G'$ has $\mathcal{O}(k^4)$ vertices, and (iii) $k' \leq k$. We also give a fixed parameter tractable (FPT) algorithm for the problem that runs in $\mathcal{O}(7^k k \cdot n^2)$ time.

## 1 Introduction

The treewidth of a graph is a measure of how "tree-like" the graph is. The notion of treewidth was introduced by Robertson and Seymour in their seminal Graph Minors series [35]. It has turned out to be very important and useful, both in the theoretical study of the properties of graphs [6, 27] and in designing graph algorithms [7, 9]. A graph has treewidth at most one if and only if it is a forest (a collection of trees), and a set of vertices in a graph $G$ whose removal from $G$ results in a forest is called a *feedback vertex set* (FVS) of the graph.

Given a graph $G$ and an integer $k$ as input, the FEEDBACK VERTEX SET problem asks whether $G$ has an FVS of size at most $k$. This is one of the first problems that Karp showed to be NP-complete [25]. The problem and its variants have extensively been investigated from the point of view of various algorithmic paradigms, including approximation and parameterized algorithms. The problem is known to have a 2-factor approximation algorithm [4], and the problem parameterized by the solution size $k$ is fixed parameter tractable (FPT) and has a polynomial kernel[1].

The quest for fast FPT algorithms and small kernels for the parameterized FEEDBACK VERTEX SET problem presents an illuminative case study of the evolution of the field of fixed parameter tractability, and stands out among the many success stories of this algorithmic approach towards solving hard problems. The first FPT algorithm for the problem, with a running time of $\mathcal{O}^*(k^4!)$, was developed by Bodlaender [5] and by Downey and Fellows [18]. After a series of improvements [19, 24, 33], a running time of the form $\mathcal{O}^*(c^k)$ was first obtained by Guo et.al [22], whose algorithm ran in $\mathcal{O}^*(37.7^k)$ time. This was improved by Dehne et.al [15] to $\mathcal{O}^*(10.6^k)$ in 2007, and to the current best $\mathcal{O}^*(3.83^k)$ by Cao et.al [13] in 2010. For classes of graphs that exclude a fixed minor $H$ (for example, planar graphs), Dorn et.al [17] have recently obtained an FPT algorithm for the problem with a running time of the form $\mathcal{O}^*(2^{O(\sqrt{k})})$.

---

[*]Accepted at WG 2010 (http://wg2010.thilikos.info/).
[1]See Section 2 for the terminology and notation used in this paper.

Proving polynomial bounds on the size of the kernel for different parameterized problems has been a significant practical aspect in the study of the parameterized complexity of NP-hard problems, and many positive results are known. See [23] for a survey of kernelization results. The existence of a polynomial kernel for the FEEDBACK VERTEX SET problem was open for a long time. It was settled in the affirmative by Burrage et. al [12] as recently as 2006, when they exhibited a kernel with $\mathcal{O}(k^{11})$ vertices. This was soon improved to a cubic vertex-kernel ($\mathcal{O}(k^3)$ vertices) by Bodlaender [8, 10]. The current smallest kernel, on $\mathcal{O}(k^2)$ vertices, is due to Thomassé [36].

The *pathwidth* of a graph is a notion closely related to treewidth, and was also introduced by Robertson and Seymour in the Graph Minors series [34]. The pathwidth of a graph denotes how "path-like" it is. A graph has pathwidth at most one if and only if it is a collection of *caterpillars*, where a caterpillar is a special kind of tree: it is a tree that becomes a path (called the *spine* of the caterpillar) when all its pendant vertices are removed. Graphs of pathwidth at most one are thus a very special kind of forests, and have even less structure than forests (which are themselves very "simple" graphs). As a consequence, some problems that are NP-hard even on forests can be solved in polynomial time on graphs of pathwidth at most one. Examples include (Weighted) Bandwidth [3, 32, 28], the Proper Interval Colored Graph problem, and the Proper Colored Layout problem [1].

In contrast to the case of forests, the corresponding vertex deletion problem for obtaining a collection of caterpillars (equivalently, a graph of pathwidth at most one) has not received much attention in the literature. In fact, to the best of our knowledge, the following problem has not yet been investigated at all: Given a graph $G$ and an integer $k$ as input, find whether $G$ contains a set of at most $k$ vertices whose removal from $G$ results in a graph of pathwidth at most one. We call such a set of vertices a pathwidth-one deletion set (PODS), and the problem the PATHWIDTH-ONE VERTEX DELETION problem. It follows from a general NP-hardness result of Lewis and Yannakakis that this problem is NP-complete.

*Our results*. We study the parameterized complexity of the PATHWIDTH-ONE VERTEX DELETION problem parameterized by the solution size $k$, and show that (i) the problem has a vertex-kernel of size $\mathcal{O}(k^4)$, and (ii) the problem can be solved in $\mathcal{O}^*(7^k)$ time (Compare with the values $\mathcal{O}(k^2)$ and $\mathcal{O}^*(3.83^k)$ for FVS, respectively).

Note that, in general, a PODS "does more" than an FVS: It "kills" all cycles in the graph, like an FVS, and, in addition, it kills all non-caterpillar trees in the graph. In fact, the difference in the sizes of a smallest FVS and a smallest PODS of a graph can be arbitrarily large. For example, the treewidth of a binary tree is one, while for any integer $c$ there exists a binary tree $T_c$ of pathwidth at least $c + 1$. Removing a single vertex from a graph will reduce the pathwidth by at most one, and so for $T_c$, the difference between the two numbers is at least $c$. Partly as a consequence of such differences, many of the techniques and reduction rules that have been developed for obtaining FPT algorithms and kernels for the FEEDBACK VERTEX SET problem do *not* carry over to the PATHWIDTH-ONE VERTEX DELETION problem. Instead, we use a characterization of graphs of pathwidth at most one to obtain the FPT algorithm and the polynomial kernel.

*Update*. After this paper was presented at WG 2010, Cygan et. al [14] improved both the results in the paper. Using the same general idea of our FPT algorithm and a clever branching strategy, they obtained an $O^*(4.65^k)$ FPT algorithm for the problem. Using some of our reduction rules and a different approach based on the $\alpha$-expansion Lemma of Thomassé [36], they obtained a quadratic ($O(k^2)$) kernel as well.

*Organization of the rest of the paper*. In Section 2 we give an overview of the notation and terminology used in the rest of the paper. In Section 3 we formally define the PATHWIDTH-ONE VERTEX DELETION problem, show that the problem is NP-complete, and sketch an FPT algorithm for the problem that runs in $\mathcal{O}^*(7^k)$ time. We show in Section 4 that the problem has a vertex-kernel of size $\mathcal{O}(k^4)$. We conclude in Section 5.

## 2 Preliminaries

In this section we state some definitions related to graph theory and parameterized complexity, and give an overview of the notation used in this paper; we also formally define the PATHWIDTH-ONE VERTEX DELETION problem and show that it is NP-hard. In general we follow the graph terminology of [16]. For a vertex $v \in V$ in a graph $G = (V, E)$, we call the set $N(v) = \{u \in V | (u, v) \in E\}$ the *open neighborhood* of $v$. The elements of $N(v)$ are said to be the *neighbors* of $v$, and $N[v] = N(v) \cup \{v\}$ is called the *closed neighborhood* of $v$. For a set of vertices $X \subseteq V$, the open and closed neighborhoods of $X$ are defined, respectively, as $N(X) = \bigcup_{u \in X} N(u) \setminus X$ and $N[X] = N(X) \cup X$. For vertices $u, v$ in $G$, $u$ is said to be a *pendant* vertex of $v$ if $N(u) = \{v\}$. A *caterpillar* is a tree that becomes a path (called the *spine* of the caterpillar) when all its pendant vertices are removed. A nontrivial caterpillar is one that contains at least two vertices. A $T_2$ is the graph on seven vertices shown in Figure 1. The *center* of a $T_2$ is the one vertex of degree 3, and its *leaves* are the three vertices of degree 1.

The operation of *contracting* an edge $(u, v)$ consists of deleting vertex $u$, renaming vertex $v$ to $uv$, and adding a new edge $(x, uv)$ for each edge $(x, u); x \neq v$. Multiple edges that may possibly result from this operation are preserved. Note that the operation is symmetric with respect to $u$ and $v$. A graph $H$ is said to be a *minor* of a graph $G$ if a graph isomorphic to $H$ can be obtained by contracting zero or more edges of some subgraph of $G$.

A *graph property* is a subset of the set of all graphs. Graph property $\Pi$ is said to hold for graph $G$ if $G \in \Pi$. $\Pi$ is said to be nontrivial if $\Pi$ and its complement are both infinite. $\Pi$ is said to be *hereditary* if $\Pi$ holds for every induced subgraph of graph $G$ whenever it holds for $G$. The *membership testing* problem for $\Pi$ is to test whether $\Pi$ holds for a given input graph.

A *tree decomposition* of a graph $G = (V, E)$ is a pair $(T, \chi)$ in which $T = (V_T, E_T)$ is a tree and $\chi = \{\chi_i \mid i \in V_T\}$ is a family of subsets of $V$, called *bags*, such that

(i) $\bigcup_{i \in V_T} \chi_i = V$;

(ii) for each edge $(u, v) \in E$ there exists an $i \in V_T$ such that both $u$ and $v$ belong to $\chi_i$; and

(iii) for all $v \in V$, the set of nodes $\{i \in V_T \mid v \in \chi_i\}$ induces a connected subgraph of $T$.

The maximum of $|\chi_i| - 1$, over all $i \in V_T$, is called the *width* of the tree decomposition. The *treewidth* of a graph $G$ is the minimum width taken over all tree decompositions of $G$. A *path decomposition* of a graph $G = (V, E)$ is a tree decomposition of $G$ where the underlying tree $T$ is a path. The *pathwidth* of $G$ is the minimum width over all possible path decompositions of $G$.

To describe the running times of algorithms we sometimes use the $\mathcal{O}^*$ notation. Given $f : \mathbb{N} \to \mathbb{N}$, we define $\mathcal{O}^*(f(n))$ to be $O(f(n) \cdot p(n))$, where $p(\cdot)$ is some polynomial function. That is, the $\mathcal{O}^*$ notation suppresses polynomial factors in the expression for the running time.

A parameterized problem $\Pi$ is a subset of $\Sigma^* \times \mathbb{N}$, where $\Sigma$ is a finite alphabet. An instance of a parameterized problem is a tuple $(x, k)$, where $k$ is called the parameter. A central notion in parameterized complexity is *fixed-parameter tractability (FPT)* which means, for a given instance $(x, k)$, decidability in time $f(k) \cdot p(|x|)$, where $f$ is an arbitrary function of $k$ and $p$ is a polynomial. The notion of *kernelization* is formally defined as follows.

**Definition 1.** [**Kernelization, Kernel**] [21, 31] A kernelization algorithm for a parameterized problem $\Pi \subseteq \Sigma^* \times \mathbb{N}$ is an algorithm that, given $(x, k) \in \Sigma^* \times \mathbb{N}$, outputs, in time polynomial in $|x| + k$, a pair $(x', k') \in \Sigma^* \times \mathbb{N}$ such that (a) $(x, k) \in \Pi$ if and only if $(x', k') \in \Pi$ and (b) $|x'|, k' \leq g(k)$, where $g$ is some computable function. The output instance $x'$ is called the kernel, and the function $g$ is referred to as the size of the kernel. If $g(k) = k^{O(1)}$ then we say that $\Pi$ admits a polynomial kernel.

When a kernelization algorithm outputs a graph on $h(k)$ vertices, we sometimes say that the output is an $h(k)$ *vertex-kernel*.

## 3 The PATHWIDTH-ONE VERTEX DELETION problem

In this section we formally define the PATHWIDTH-ONE VERTEX DELETION problem, show that it is NP-complete, and briefly sketch an $\mathcal{O}^*(7^k)$ FPT algorithm for the problem. We begin with the observation that caterpillars are the quintessential graphs of pathwidth at most one:

**Fact 1.** [2] *A graph $G$ has pathwidth at most one if and only if it is a collection of vertex-disjoint caterpillars.*

A vertex set $S \subseteq V$ of a graph $G$ is said to be a *pathwidth-one deletion set* (PODS) if $G[V \setminus S]$ has pathwidth at most one. In this paper we investigate the parameterized complexity of the following problem:

PATHWIDTH-ONE VERTEX DELETION (POVD)
*Input:* An undirected graph $G = (V, E)$, and a positive integer $k$.
*Parameter:* $k$
*Question:* Does there exist a set $S \subseteq V$ of at most $k$ vertices of $G$ such that $G[V \setminus S]$ has pathwidth at most one (i.e., $S$ is a PODS of $G$)?

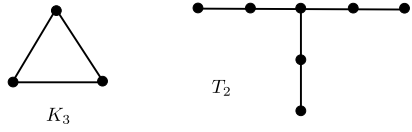

Figure 1: The set of excluded minors for graphs of pathwidth at most one.

The following general NP-completeness result is due to Lewis and Yannakakis:

**Fact 2.** [29] *The following problem is* NP-*complete for any nontrivial hereditary graph property* Π *for which the membership testing problem can be solved in polynomial time:*

Input:   *Graph $G = (V, E)$, positive integer $k$.*
Question:  *Is there a subset $S \subseteq V, |S| \leq k$ such that $G[V \setminus S] \in \Pi$?*

The NP-completeness of the PATHWIDTH-ONE VERTEX DELETION problem follows directly from this result:

**Theorem 1.** [⋆][2] *The* PATHWIDTH-ONE VERTEX DELETION *problem is* NP-*complete.*

In the rest of the paper we focus on the parameterized complexity of the PATHWIDTH-ONE VERTEX DELETION problem. We now sketch an $\mathcal{O}^*(7^k)$ time FPT algorithm, and in the next section we describe an $\mathcal{O}(k^4)$ vertex-kernel for the problem. Let $(G = (V, E), k)$ be the input instance, where $|V| = n$. Let $S \subseteq V$ be a PODS of $G$ of size at most $k$. Observe that if $(G, k)$ is a YES instance, then the number of edges in $G$ is at most $k(n-1) + (n-1) = (k+1)(n-1)$. The first term on the left is an upper bound on the number of edges that are incident the vertices in $S$; the second term is a loose upper bound on the number of edges in $G \setminus S$. So, if $G$ has more than $(k+1)(n-1)$ edges, then we can immediately reject the input. Since each reduction rule in the sequel is sound, and no rule increases the number of vertices or edges, from now on we assume, without loss of generality, that the graph has at most $(k+1)(n-1)$ edges.

The kernel arguments are based on Fact 1, while our starting point for the FPT algorithm is the following characterization, in terms of excluded minors, of graphs of pathwidth at most one:

**Fact 3.** [11, 20] *A graph $G$ has pathwidth at most one if and only if it does not contain $K_3$ or $T_2$ as a minor, where $K_3$ and $T_2$ are as in Figure 1.*

Fact 3 is not very helpful in the given form in checking for a small PODS. Instead, we derive and use the following alternate characterization and the two succeeding lemmas:

**Lemma 1.** [⋆] *A graph $G$ has pathwidth at most one if and only if it does not contain a cycle or a $T_2$ as a subgraph.*

**Lemma 2.** [⋆] *Let $\mathcal{S} = \{T_2, K_3, C_4\}$, where $C_4$ is a cycle of length $4$. Given a graph $G = (V, E); |V| = n$, we can find whether $G$ contains a subgraph $H$ that is isomorphic to one of the graphs in $\mathcal{S}$, and also locate such an $H$ if it exists, in $\mathcal{O}(kn^2)$ time.*

**Lemma 3.** [⋆] *Let $\mathcal{S} = \{T_2, K_3, C_4\}$, where $C_4$ is a cycle of length $4$. If $G$ is a graph that does not contain any element of $\mathcal{S}$ as a subgraph, then each connected component of $G$ is either a tree, or a cycle with zero or more pendant vertices ("hairs") attached to it.*

---

[2]Proofs of results labeled with a [⋆] have been moved to the Appendix due to space constraints.

## 3.1 An FPT algorithm for POVD

Let $(G = (V, E), k)$ be the input instance, where $|V| = n$. We use a branching strategy inspired by Lemmas 1 and 3. First we locate a (not necessarily induced) subgraph $T$ of $G$ that is isomorphic to one of $\mathcal{S} = \{T_2, K_3, C_4\}$. From Lemma 2, this can be done in $\mathcal{O}(kn^2)$ time. At least one of the (at most seven) vertices of $T$ must be in any PODS of $G$. So we branch on the vertices of $T$: We pick each one, in turn, into the minimal PODS that we are constructing, delete the picked vertex and all its adjacent edges, and recurse on the remaining graph after decrementing the parameter by one.

The leaves of this recursion tree correspond to graphs which do not have a subgraph isomorphic to any graph in $\mathcal{S}$. By Lemma 3, each connected component of such a graph is a tree, or a cycle with zero or more pendant vertices ("hairs") attached to it. The trees can be ignored — they do not have a $T_2$ as a subgraph — and each cycle (with or without hairs) forces exactly one vertex into any minimal solution. Thus the base case of the recursion can be solved in linear time.

This is a 7-way branching, where the depth of the recursion is at most $k$, and where the algorithm spends $\mathcal{O}(kn^2)$ time at each node. Hence we have

**Theorem 2.** *The* PATHWIDTH-ONE VERTEX DELETION *problem parameterized by the solution size $k$ has an FPT algorithm that runs in $\mathcal{O}(n^2 \cdot 7^k k)$ time.*

By a folklore result of parameterized complexity, it follows immediately from Theorem 2 that the PATHWIDTH-ONE VERTEX DELETION problem parameterized by the solution size $k$ has a kernel of size $\mathcal{O}(7^k)$ (See, for example, [21]). We now show that the kernel size can be brought down significantly from this trivial bound.

## 4 A polynomial kernel for POVD

We turn to the main result of this paper. We describe a polynomial-time algorithm (the *kernelization algorithm*) that, given an instance $(G, k)$ of POVD, returns an instance $(G', k')$ (the *kernel*) of POVD such that (i) $(G, k)$ is a YES instance if and only if $(G', k')$ is a YES instance, (ii) $G'$ has $O(k^4)$ vertices, and (iii) $k' \leq k$. The kernelization algorithm (Algorithm 1) exhaustively applies the reduction rules of Section 4.1 to the input instance. The resulting instance, to which no rule applies, is said to be *reduced* with respect to the reduction rules. To demonstrate a quartic vertex-kernel for the problem, it suffices to show that

1. The rules can be exhaustively applied in polynomial time;

2. Each rule is *sound*: the output of a rule is a YES instance if and only if its input is a YES instance; and

3. If the input instance $(G, k)$ is a YES instance, then the reduced instance $(G', k')$ has $O(k^4)$ vertices.

The reduction rules are based on the following idea: Suppose $(G = (V, E), k)$ is a YES instance of the problem that is reduced with respect to the reduction rules. Then there is a set $S \subseteq V, |S| \leq k$ such that $G[V \setminus S]$ is a collection of caterpillars, and it suffices to show that $|V \setminus S| = O(k^4)$. We express $V \setminus S$ as the union of different kinds of vertices, and devise reduction rules that help us bound the total number of vertices of each kind. To be more specific, we set $V \setminus S = V_1 \cup V_2 \cup V_3 \cup V_4 \cup V_5$ where

**Algorithm 1** The kernelization algorithm
1: **procedure** KERNELIZE($G, k$)
2:     $CurrentInstance \leftarrow (G, k)$
3:     **repeat**
4:         Apply Rules 1 to 6, updating $CurrentInstance$ with the output of each rule.
5:     **until** None of the rules cause any change to $CurrentInstance$.
6: **end procedure**

1. $V_1 = \{v \in (V \setminus S); N(v) \cap (V \setminus S) = \emptyset \text{ and } |N(v) \cap S| \leq 1\}$
2. $V_2 = \{v \in (V \setminus S); N(v) \cap (V \setminus S) = \emptyset \text{ and } |N(v) \cap S| \geq 2\}$
3. $V_3 = \{v \in ((V \setminus S) \setminus V_1); v \text{ lies on the spine of a nontrivial caterpillar in } G[V \setminus S]\}$
4. $V_4 = \{v \in (V \setminus S); |N(v) \cap S| = 0 \text{ and } v \text{ is a pendant vertex in } G[V \setminus S]\}$
5. $V_5 = \{v \in (V \setminus S); |N(v) \cap S| \geq 1 \text{ and } v \text{ is a pendant vertex in } G[V \setminus S]\}$

It is easy to verify that these sets together exhaust $V \setminus S$. We state the reduction rules and describe their consequences in the next section; the proofs of soundness of the rules and a more formal bound on the running time and kernel size are deferred to Section 4.2.

### 4.1 Reduction Rules

For each rule below, let $(H = (V_H, E_H), k)$ be the instance on which the rule is applied, and $(H', k')$ the resulting instance. Let $G = (V, E)$ be a YES instance of the problem that is reduced with respect to all the reduction rules, and let $S, V_1, \ldots, V_5$ be as described above. To bound the sizes of various subsets of $V \setminus S$, we use the fact that no reduction rule applies to $G$.

**Rule 1.** *If a connected component $H[X]; X \subseteq V_H$ of $H$ has pathwidth at most 1, then remove $X$ from $H$. The resulting instance is $(H' = H[V_H \setminus X], k' = k)$.*

**Rule 2.** *If a vertex $u$ in $H$ has two or more pendant neighbors, then delete all but one of these pendant neighbors to obtain $H'$. The resulting instance is $(H', k' = k)$.*

Rules 1 and 2 together ensure that every caterpillar in $G[V \setminus S]$ has at least one neighbor in $S$, and that $|V_1| \leq k$: See Lemma 5.

**Rule 3.** *Let $u$ be a vertex of $H$ with at least two neighbors. If for every two vertices $\{v, w\} \subseteq N(u)$ there exist $k + 2$ vertices excluding $u$ that are adjacent to both $v$ and $w$, then delete $u$ from $H$. The resulting instance is $(H' = H[V_H \setminus \{u\}], k' = k)$.*

Rule 3 ensures that $|V_2| \leq \binom{k}{2}(k+2)$: Set $A = V_2$ and $X = S$ in Lemma 6.

**Rule 4.** *For a vertex $u$ of $H$, if there is a matching $M$ of size $k+3$ in $H$ where (i) each edge in $M$ has at least one end vertex in $N(u)$, and, (ii) $u$ is not incident with any edge in $M$, then delete $u$ and decrement $k$ by one. The resulting instance is $(H' = H[V \setminus \{u\}], k' = k - 1)$.*

**Rule 5.** *Let $x, y$ be the end vertices of the spine $x, v_1, v_2, v_3 \ldots, v_p, y$ of an induced caterpillar $C$ in $H$ such that (1) no $v_i; 1 \leq i \leq p$ is adjacent in $H$ to any vertex outside $C$, and (2) every pendant vertex of $C$ is a pendant vertex in $H$. If $p \geq 5$, then contract the edge $(v_2, v_3)$ in $H$ to obtain the graph $H'$. The resulting instance is $(H', k = k')$.*

From Rules 1 to 5 it follows that $|V_3| \leq 17k(k+2)$ (Lemma 7), and that $|V_5| \leq 17(k+2)^2k(2k-1)$ (Lemma 8). Each vertex in $G$ can have at most one pendant neighbor, or else Rule 2 would apply. From this we get $|V_4| \leq |V_3| = 17k(k+2)$. Putting all the bounds together, $|V| \leq 34k^4 + 120k^3 + 103k^2 + k$, and so we have:

**Rule 6.** *If none of the Rules 1 to 5 can be applied to the instance $(H, k)$, and $|V_H| > 34k^4 + 120k^3 + 103k^2 + k$ , then set the resulting instance to be the trivial NO instance $(H', k')$ where $H'$ is a cycle of length $3$ and $k' = 0$.*

In the next section we prove that these rules are sound, and that they can all be applied exhaustively in polynomial time. Hence we get

**Theorem 3.** *The* PATHWIDTH-ONE VERTEX DELETION *problem parameterized by solution size $k$ has a polynomial vertex-kernel on $\mathcal{O}(k^4)$ vertices.*

### 4.2 Correctness and running time

We now show that the reduction rules are sound. That is, we show that for each rule, (using the notation of the previous section) $(H, k)$ is a YES instance if and only if $(H', k')$ is a YES instance. We also show that each rule can be implemented in polynomial time. For discussing the rules, we reuse the notation from the respective rule statement in Section 4.1. In each case, $n$ is the number of vertices in the input to the kernelization algorithm. For want of space, we have moved all the proofs to the Appendix.

*Claim* 1. [⋆] Rule 1 is sound, and can be applied in $\mathcal{O}(kn)$ time.

*Claim* 2. [⋆] Rule 2 is sound, and can be applied in $\mathcal{O}(kn)$ time.

*Claim* 3. [⋆] Rule 3 is sound, and can be applied in $\mathcal{O}(n^3)$ time.

*Claim* 4. [⋆] Rule 4 is sound, and can be applied in $\mathcal{O}(kn^{1.5})$ time.

*Claim* 5. [⋆] Rule 5 is sound, and can be applied in $\mathcal{O}(kn)$ time.

From these claims, we get

**Lemma 4.** *On an input instance $(G = (V, E), k); |V| = n$ of* PATHWIDTH-ONE VERTEX DELETION*, the kernelization algorithm (Algorithm 1) runs in $\mathcal{O}(n^4)$ time and outputs a kernel on $\mathcal{O}(k^4)$ vertices.*

*Proof.* From Claims 1 to 5 it follows that Rules 1 to 5 are sound, and that each can be applied in $\mathcal{O}(n^3)$ time. From the discussion in Section 4.1 (using Lemmas 5 to 8 below) it follows that Rule 6 is sound, and it is easy to see that this rule can be applied in $\mathcal{O}(n)$ time. Each time a rule is applied, the number of vertices in the graph reduces by at least one (contracting an edge also reduces the vertex count by one). Hence the loop in lines 3 to 5 of Algorithm 1 will run at most $|V| + 1 = n + 1$ times. The algorithm produces its output either at a step where Rule 6 applies, or when none of the rules applies and the remaining instance has $\mathcal{O}(k^4)$ vertices. Thus the algorithm runs in $\mathcal{O}(n^4)$ time and outputs a kernel on $\mathcal{O}(k^4)$ vertices. □

The remaining lemmas in this section are used in in Section 4.1 to bound the sizes of the sets $V_1, \ldots, V_5$. Their proofs are deferred to the Appendix.

**Lemma 5.** [⋆] *Let $(G = (V, E), k)$ be a YES instance of the problem that is reduced with respect to Rules 1 and 2, and let $S$ be a PODS of $G$ of size at most $k$. Let $V_1 = \{v \in (V \setminus S); (N(v) \cap (V \setminus S)) = \emptyset$ and $|N(v) \cap S| \leq 1\}$. Then every caterpillar in $G[V \setminus S]$ has at least one neighbor in $S$, and $|V_1| \leq k$.*

**Lemma 6.** [⋆] *Let $(G, k)$ be a YES instance of the problem that is reduced with respect to Rule 3. For a set $X \subseteq V$, if $A \subseteq V \setminus X$ is such that every $v \in A$ has (i) at least two neighbors in $X$, and (ii) no neighbors outside $X$, then $|A| \leq \binom{|X|}{2}(k+2)$.*

**Lemma 7.** [⋆] *Let $(G = (V, E), k)$ be an instance of the problem that is reduced with respect to Rules 1 to 5, and let $S \subseteq V$ be such that $G[V \setminus S]$ has pathwidth at most one. Let $X \subseteq (V \setminus S)$ be the set of vertices in $(V \setminus S)$ that lie on the spines of nontrivial caterpillars in $G[V \setminus S]$. Then $|X| \leq 17k(k+2)$.*

**Lemma 8.** [⋆] *Let $(G = (V, E), k)$ be a YES instance of the problem that is reduced with respect to Rules 1 to 5, and let $S \subseteq V; |S| \leq k$ be such that $G[V \setminus S]$ has pathwidth at most one. Let $P \subseteq (V \setminus S)$ be the set of pendant vertices in $G[V \setminus S]$ that have at least one neighbor in $S$. Then $|P| \leq 17(k+2)^2 k(2k-1)$.*

## 5 Conclusion

We defined the PATHWIDTH-ONE VERTEX DELETION problem as a natural variant of the iconic FEEDBACK VERTEX SET problem, and initiated the study of its algorithmic complexity. We established that the problem is NP-complete, and showed that the problem parameterized by the solution size $k$ is fixed-parameter tractable. We gave an FPT algorithm for the problem that runs in $\mathcal{O}^*(7^k)$ time, and showed that the problem has a polynomial kernel on $\mathcal{O}(k^4)$ vertices.

An immediate question is whether these bounds can be improved upon. A more challenging problem is to try to solve the analogous problem for larger values of pathwidth. That is, we know that for any positive integer $c$, the Pathwidth $c$ Vertex Deletion problem, defined analogously to PATHWIDTH-ONE VERTEX DELETION, is FPT parameterized by the solution size. This follows from the Graph Minor Theorem of Robertson and Seymour because, for each fixed $c$, the set of YES instances for this problem form a minor-closed class. However, for $c = 2$, the number of graphs in the obstruction set is already a hundred and ten [26], and so our approach would probably be of limited use for $c \geq 2$. Thus the interesting open problems for $c \geq 2$ are: (i) Can we get an $\mathcal{O}^*(d^k)$ FPT algorithm for the problem for some constant $d$, and (ii) Does the problem have a polynomial kernel?

**Acknowledgements.** We thank our anonymous reviewers for a number of useful comments for improving the paper.

# A  Deferred Proofs

In this appendix we state those proofs that were omitted from the main body of the paper due to space constraints.

## A.1  The PATHWIDTH-ONE VERTEX DELETION problem

**Theorem 1.** *The* PATHWIDTH-ONE VERTEX DELETION *problem is* NP-*complete.*

*Proof.* Let $\Pi$ be the set of all graphs of pathwidth at most one. It is easy to verify that $\Pi$ is a nontrivial hereditary graph property. By a direct application of the definition of a caterpillar, we can check in polynomial time whether each component of a graph $G$ is a caterpillar. Together with Fact 1, this implies that membership testing for $\Pi$ can be done in polynomial time. The theorem now follows from Fact 2. □

**Lemma 1.** *A graph $G$ has pathwidth at most one if and only if it does not contain a cycle or a $T_2$ as a subgraph.*

*Proof.* If $G$ has pathwidth at most one, then by Fact 3 it does not contain $K_3$ or $T_2$ as a minor; it follows that $G$ does not contain any cycle or $T_2$ as a subgraph.

Conversely, assume that $G$ does not contain a cycle or a $T_2$ as a subgraph. Suppose $G$ contains a $K_3$, say $C$, as a minor, obtained by contracting the set of edges $E'$ and deleting the set of edges $E''$ of a subgraph $H$ of $G$. If we replace the contracted vertices with the original edges $E'$ and add the deleted edges $E''$, we obtain a cycle which is a subgraph of the original graph (from which we obtained $C$ as a minor), a contradiction.

Similarly, if $G$ has a $T_2$ as a minor, then replacing the contracted vertices with the original edges and adding the deleted edges gives rise to a supergraph of $T_2$, a contradiction. Thus $G$ does not contain either $K_3$ or $T_2$ as a minor, and so by Fact 3 $G$ has pathwidth at most one. □

**Lemma 2.** *Let $\mathcal{S} = \{T_2, K_3, C_4\}$, where $C_4$ is a cycle of length $4$. Given a graph $G = (V, E); |V| = n$, we can find whether $G$ contains a subgraph $H$ that is isomorphic to one of the graphs in $\mathcal{S}$, and also locate such an $H$ if it exists, in $\mathcal{O}(kn^2)$ time.*

*Proof.* It is well-known that we can find the girth (length of a shortest cycle) of a graph by doing a breadth-first search (BFS) from each vertex. The same algorithm finds a smallest cycle in the graph as well, and so we can use it to check for and locate a $K_3$ or $C_4$ in $G$. Suppose $G$ contains neither of these graphs as a subgraph. To check if $G$ contains a $T_2$, we guess the center vertex $v$ of the $T_2$ and do a BFS starting from $v$ (the level 0 vertex). Since $G$ does not contain $K_3$ as a subgraph, there is a $T_2$ with $v$ as the center if and only if at least three vertices in level 1 of the BFS have at least one neighbor each in level 2. We can combine the two tests to obtain an algorithm of the required kind that runs in $O((|V|)(|V|+|E|)) = O(n(n+(k+1)(n-1))) = O(kn^2)$ time. □

**Lemma 3.** *Let $\mathcal{S} = \{T_2, K_3, C_4\}$, where $C_4$ is a cycle of length $i$. If $G$ is a graph that does not contain any element of $\mathcal{S}$ as a subgraph, then each connected component of $G$ is either a tree, or a cycle with zero or more pendant vertices ("hairs") attached to it.*

*Proof.* Let $G$ be a graph that does not contain any element of $\mathcal{S}$ as a subgraph, and let $X$ be a connected component of $G$ that is not a tree. Then $X$ contains a cycle. Let $C$ be a smallest cycle in $X$. Then $C$ has length at least $5$. Suppose there is a path $<a, b, c>$ in $X$, where $a$ is a vertex on $C$ and $b$ is not. Then $c \notin V(C)$, or else $a, b, c$ and the shorter path from $c$ to $a$ on $C$ form a cycle of length at most $4$, a contradiction. Let $x, y$ be the two neighbors of $a$ on $C$, and let $x' \neq a, y' \neq a$ be neighbors of $x, y$ on $C$. Then $x, y, x', y'$ are all distinct, and $a, b, c, x, y, x', y'$ form a $T_2$ in $X$ with $a$ at the center, a contradiction. It follows that for any vertex $u \in V(C)$, any neighbor $v \notin V(C)$ of $u$ is a pendant vertex in $X$, and the lemma follows from this. □

## A.2 Correctness and running time of the kernelization algorithm

In the proofs for a rule, we reuse the notation from the respective rule statement in Section 4.1. In each case, $n$ is the number of vertices in the input to the kernelization algorithm.

*Claim 1.* Rule 1 is sound, and can be applied in $\mathcal{O}(kn)$ time.

*Proof.* The connected component $H[X]$ does not intersect any forbidden structure and thus does not affect any solution of the problem.

By doing a single breadth-first search (BFS) of $H$, we can find all the acyclic connected components of $H$. To check if an acyclic component $C$ contains a $T_2$, we delete all the leaves (vertices of degree one) in $C$ and check if the remaining graph is a simple path. $C$ contains a $T_2$ if and only if the remaining graph is *not* a simple path. Using a queue and an array to keep track of the leaves and the degrees of the vertices in $C$, respectively, all this can be done in $O((|V_H| + |E_H|)) = O(n + (k+1)(n-1)) = O(kn)$ time. □

**Observation 1.** *Let $G$ be any graph of pathwidth at most one. Adding new degree zero vertices to $G$ or adding new pendant neighbors to an isolated vertex $u$ of $G$ does not add a cycle or a $T_2$ to $G$.*

**Lemma 9.** *Let $G$ be any graph that does not have any subgraph isomorphic to $T_2$. If $v$ is a vertex in $G$ such that by adding some number $l \geq 1$ of pendant vertices as neighbors to $v$ we can obtain a graph $H$ that contains a $T_2$ as a subgraph, then $v$ has no pendant neighbors in $G$.*

*Proof.* Assume to the contrary that $v$ has a pendant neighbor $u$ in $G$. Consider any $T_2$, say $t$, in $H$. $t$ contains at least one of the new pendant vertices; say it contains $w; w \neq u$. Since $w$ is pendant in $H$, the degree of $w$ in $t$ is exactly $1$, and so $w$ is one of the leaf vertices of $t$; its only neighbor in $t$ is $v$, which in turn is a non-central internal vertex of $t$. None of the internal vertices of a $T_2$ has two distinct pendant neighbors, and so $u$ is not in $t$. It is evident that one can remove $w$ from $t$ and add $u$ and the edge $\{u, v\}$ to the resulting subgraph to obtain a $T_2$ consisting entirely of vertices in $G$, a contradiction. □

*Claim 2.* Rule 2 is sound, and can be applied in $\mathcal{O}(kn)$ time.

*Proof. Soundness.* Let $L$ be the set of pendant neighbours of $u$ in $H$ that are deleted to obtain the graph $H'$, and let $r$ be the pendant neighbour of $u$ remaining in $H'$. Let $H = (V_H, E_H), H' = (V'_H, E'_H)$, so that $V'_H = V_H \setminus L$ and $H' = H[V_H \setminus L]$. We have to show that

> There exists a set $S \subseteq V_H, |S| \leq k$ such that $H[V_H \setminus S]$ contains no cycles or $T_2$s (i.e, has pathwidth one) if and only if there exists a set $S' \subseteq V'_H, |S'| \leq k$ such that $H'[V'_H \setminus S']$ contains no cycles or $T_2$s.

($\Longrightarrow$): If there exists an $S \subseteq V_H, |S| \leq k$ such that $H[V_H \setminus S]$ contains no cycles or $T_2$s, then let $S' = S \setminus L$. Clearly $S' = S \setminus L \subseteq V_H \setminus L = V'_H$, and $|S'| = |S| - |S \cap L| \leq |S| \leq k$. Now $H'[V'_H \setminus S'] = H'[(V_H \setminus L) \setminus (S \setminus L)] = H[(V_H \setminus L) \setminus (S \setminus L)] = H[(V_H \setminus S) \setminus L]$, and since $H[V_H \setminus S]$ contains no cycles or $T_2$s, neither does its induced subgraph $H[(V_H \setminus S) \setminus L] = H'[V'_H \setminus S']$.

($\Longleftarrow$): If there exists a set $S' \subseteq V'_H, |S'| \leq k$ such that $H'[V'_H \setminus S']$ contains no cycles or $T_2$s, then let $K = H[V_H \setminus S']$. Now, the vertices in $L$ have degree at most one in $K$, and so do not belong to any cycle in $K$. Therefore, if $K$ contains a cycle, then so does $K \setminus L = H[V_H \setminus S' \setminus L] = H[(V_H \setminus L) \setminus S'] = H'[V'_H \setminus S']$, which contradicts the assumption that $H'[V'_H \setminus S']$ contains no cycles or $T_2$s. So $K$ does not contain a cycle.

If $K$ does not contain a $T_2$ either, then setting $S = S'$ completes the argument. So let $K$ contain a $T_2$. Then $H'[V'_H \setminus S'] = H[(V_H \setminus L) \setminus S'] = H[V_H \setminus S' \setminus L]$ contains no $T_2$, and $K = H[V_H \setminus S']$ contains a $T_2$. The vertices in $L$ have degree at most one in $K$, and from the first part of Observation 1 it follows that these vertices have degree exactly one in $K$. So $u \in K$, i.e., $u \in V_H \setminus S'$, and by the definition of $L$, $u \notin L$. Thus $u \in V_H \setminus S' \setminus L$, and by Lemma 9, $r \notin V_H \setminus S' \setminus L$. But by definition $r \notin L$, and so $r$ must be in $S'$. Now, if there is a $T_2$ $t$ in $K$ that does not contain $u$, then $t$ does not contain any vertex from $L$ either, and so $t$ is present in $K \setminus L = H[V_H \setminus S' \setminus L]$, a contradiction. So every $T_2$ in $K$ contains $u$.

Set $S := (S' \setminus \{r\}) \cup \{u\}$. Clearly $S \subseteq V_H$, and $|S| = |S'| \leq k$. Since $K = H[V_H \setminus S']$ does not contain a cycle, and since the only neighbour of $r$ in $H$ is $u$, adding $r$ to $K$ and removing $u$ does not introduce a cycle. Since every $T_2$ in $K$ contains $u$, removing $u$ from $K$ also removes all $T_2$s from $K$. Since the only neighbour of $r$ in $H$ is $u$, adding $r$ to $K$ and removing $u$ does not introduce a $T_2$. Thus $H[V_H \setminus ((S' \setminus \{r\}) \cup \{u\})] = H[V_H \setminus S]$ contains no cycles or $T_2$s.

*Running time.* It is easy to see that this rule can be applied in $O(|V_H| + |E_H|) = O(n + (k+1)(n-1)) = O(kn)$ time. □

**Lemma 10.** *Let $G$ be any graph of pathwidth at most one. If $v$ is a vertex of degree at least 2 in $G$ and $H$ is a graph obtained from $G$ by adding some number $l \geq 1$ of pendant vertices as neighbours to $v$, then $H$ also has pathwidth at most one.*

*Proof.* Let $u_1, u_2$ be two neighbors of $v$ in $G$. Assume to the contrary that $H$ has pathwidth more than one. It is clear that $H$ does not contain a cycle, and so by Lemma 1 $H$ contains a subgraph $K$ isomorphic to $T_2$. $K$ contains at least one of the new pendant vertices; say it contains $w$. Since $w$ is pendant in $H$, the degree of $w$ in $K$ is exactly 1, and so $w$ is one of the leaf vertices of $K$; its only neighbor in $K$ is $v$, which in turn is a degree two vertex of $K$. Further, one of the neighbors of $v$ in $G$ is the central vertex of $K$; say $u_1$ is the central vertex of $K$. Vertex $u_2$ is not part of $K$, or else the edges $(v, u_1), (v, u_2)$ and the path in $K$ from $u_1$ to $u_2$ would form a cycle in $G$. So we can remove $w$ from $K$ and add $u_2$ and the edge $(v, u_2)$ to the resulting subgraph to obtain a $T_2$ consisting entirely of vertices in $G$, a contradiction. □

*Claim* 3. Rule 3 is sound, and can be applied in $\mathcal{O}(n^3)$ time.

*Proof. Soundness.* Let $X$ be a set of at most $k$ vertices of $H$ whose removal results in a graph of pathwidth one. Then removing $X' = (X \setminus \{u\})$ from $H'$ results in a graph of pathwidth one, and $|X'| \leq k$.

For the other direction, let $X'$ be a set of at most $k$ vertices of $H'$ such that $H'[V'_H \setminus X'] = H[V'_H \setminus X']$ has pathwidth at most one. It is sufficient to show that removing $X'$ from $H$ results in a graph of pathwidth at most one. This, in turn, is equivalent to showing that adding $u$ (and all the edges from $u$ to $V_H \setminus X'$ in $H$) to $H[V'_H \setminus X']$ will result in a graph (which is $H[V_H \setminus X']$) with pathwidth at most one.

Now, since $H[V'_H \setminus X']$ has pathwidth at most one, $X'$ has at least one vertex in common with every cycle in $H[V'_H]$. Also, for any two vertices $\{v, w\} \subseteq N(u)$ there are $k+2$ vertex disjoint paths from $v$ to $w$ in $H[V'_H]$, and so either $v$ or $w$ has to be in $X'$. It follows that $|N(u) \setminus X'| \leq 1$.

If $N(u) \subseteq X'$, then $u$ is isolated in $H[V_H \setminus X']$, and it follows from Observation 1 that $H[V_H \setminus X'] = H[(V'_H \setminus X') \cup \{u\}]$ has pathwidth at most one. So suppose $N(u) \not\subseteq X'$, and let $v$ be the single vertex in $N(u) \setminus X'$. Now in $H'$, $v$ has at least $k+2$ neighbours excluding $u$, and so there are at least two such neighbours of $v$, say $y_1, y_2$, that are not in $X'$. Thus (i) $v$ has degree at least 2 in the graph $H[V'_H \setminus X']$ of pathwidth at most one, and (ii) $H[V_H \setminus X'] = H[(V'_H \setminus X') \cup \{u\}]$ is obtained by adding a pendant vertex adjacent to $v$ to $H[V'_H \setminus X']$, and so by Lemma 10, $H[V_H \setminus X']$ has pathwidth at most one.

*Running time.* The rule can be applied in $\mathcal{O}(n^3)$ time as follows. Construct a new graph $K = (V_K = V_H, E_K)$, where for each pair of vertices $x, y \in V_H$, add edge $(x, y)$ to $E_K$ if in the graph $H$, $|(N(x) \cap N(y)) \setminus \{x, y\}| \geq k + 3$. To see if $u \in V_H$ qualifies for deletion from $H$ as per the rule, check if $N(u)$ induces a clique in $K$. □

*Claim* 4. Rule 4 is sound, and can be applied in $\mathcal{O}(kn^{1.5})$ time.

*Proof. Soundness.* Let $X$ be a set of at most $k$ vertices of $H$ whose removal results in a graph of pathwidth at most one, and let $X' = (X \setminus \{u\})$. At least three edges of the matching $M$, say $\mathcal{E} = \{\{x_1, y_1\}, \{x_2, y_2\}, \{x_3, y_3\}\}$, survive in $H[V_H \setminus X]$. Without loss of generality, let $\{x_1, x_2, x_3\} \subseteq N(u)$. If $u \notin X$, then $\mathcal{E}$ and the edges $\{u, x_1\}, \{u, x_2\}, \{u, x_3\}$ together form a $T_2$ in $H[V_H \setminus X]$, a contradiction. Hence $u \in X$, and so $|X'| = |X| - 1$. Clearly, removing $X'$ from $H'$ results in a graph of pathwidth one, and $|X'| \leq k - 1$.

For the other direction, if $X'$ is a set of at most $k-1$ vertices of $H'$ such that $H'[V'_H \setminus X'] = H[V'_H \setminus X']$ has pathwidth at most one, then clearly $X = X' \cup \{u\}$ is a set of at most $k$ vertices of $H$ such that $H[V_H \setminus X]$ has pathwidth at most one.

*Running time.* The rule can be applied in $\mathcal{O}(kn^{1.5})$ time as follows. Let $A = N(u), B = N(A) \setminus \{u\}$ in $H$. Construct a new graph $K$ from $H[A \cup B]$ by deleting all the edges in $H[B]$. By doing two levels of a breadth-first traversal starting from $u$, this can be done in $O(|V_H| + |E_H|) = O(n + (k+1)(n-1)) = O(kn)$ time. Find a maximum matching $\mathcal{M}$ in $H$ in $O(\sqrt{|V_K|}|E_K|) = O(kn^{1.5})$ time [30]. $\mathcal{M}$ is a largest matching of the kind specified in the rule, and so we only have to check if $\mathcal{M}$ contains at least $k+3$ edges. □

We need the following observations to show that Rule 5 is sound.

**Observation 2.**

1. Let $G$ be any graph that contains at least one cycle. Any graph $G'$ obtained from $G$ by contracting an edge of $G$ also contains at least one cycle (possibly containing parallel edges).

2. Let $G$ be a graph that contains a $T_2$ as a subgraph. If $G'$ is a graph obtained from $G$ by contracting an edge $(u, v)$ where either $u$ or $v$ (or both) is not part of any $T_2$ in $G$, then $G'$ also contains a $T_2$ as a subgraph.

**Fact 4.** [27] *For any fixed non-negative integer $p$, the class of graphs of pathwidth at most $p$ is closed under the operation of taking minors.*

*Claim 5.* Rule 5 is sound, and can be applied in $\mathcal{O}(kn)$ time.

*Proof. Soundness.* Let $v_2v_3$ be the vertex resulting from the edge contraction. Let $X$ be a set of at most $k$ vertices of $H$ whose removal results in a graph $K$ of pathwidth at most one. If $\{v_2, v_3\} \cap X = \emptyset$, then the graph $K' = H'[V'_H \setminus X]$ is a minor of $K$: $K'$ can be obtained from $K$ by contracting the edge $\{v_2, v_3\}$. If $\{v_2, v_3\} \cap X \neq \emptyset$, then let $X' = (X \cup \{v_2v_3\}) \setminus \{v_2, v_3\}$. Clearly $|X'| \leq |X|$, and $K' = H'[V'_H \setminus X']$ is a subgraph of $K$: if $\{v_2, v_3\} \subseteq X$, then $K'$ is isomorphic to $K$, and if exactly one of $v_2, v_3$ is in $X$, then $K'$ can be obtained from $K$ by deleting the other vertex. In both cases, by Fact 4, $K'$ has pathwidth at most one, and so in all cases there is a vertex set of size at most $k$ in $H'$ whose removal gives a graph of pathwidth at most one.

For the other direction, suppose $X'$ is a minimal set of at most $k$ vertices of $H'$ such that $K' = H'[V'_H \setminus X']$ has pathwidth at most one. If $v_2v_3 \notin X'$, then $X' \subseteq V_H$, and $K'$ can be obtained from $K = H[V_H \setminus X']$ by contracting the edge $\{v_2, v_3\}$. By the contrapositive of Observation 2, $K$ contains neither a cycle nor a $T_2$. Hence $X'$ is a set of at most $k$ vertices of $H$ such that $H[V_H \setminus X']$ has pathwidth at most one. If $v_2v_3 \in X'$, then it is easy to see that $X = (X' \setminus \{v_2v_3\}) \cup \{v_2\}$ is a set of at most $k$ vertices of $H$ such that $H[V_H \setminus X']$ has pathwidth at most one.

*Running time.* The rule can be applied in $\mathcal{O}(kn)$ time as follows: first we delete all pendant vertices in the graph. This can be done in $O(|V_H|+|E_H|)$ time. In the remaining graph, we check if there is a path of length 5 or more consisting of vertices of degree two. This can be done, by doing a BFS, in $O(|V_H| + |E_H|)$ time. The total running time is thus $O(|V_H| + |E_H|) = O(kn)$. □

**Lemma 5.** *Let $(G = (V, E), k)$ be a YES instance of the problem that is reduced with respect to Rules 1 and 2, and let $S$ be a PODS of $G$ of size at most $k$. Let $V_1 = \{v \in (V \setminus S) ; (N(v) \cap (V \setminus S)) = \emptyset$ and $|N(v) \cap S| \leq 1\}$. Then every caterpillar in $G[V \setminus S]$ has at least one neighbor in $S$, and $|V_1| \leq k$.*

*Proof.* If a caterpillar in $G[V \setminus S]$ has no neighbor in $S$, then Rule 1 would apply to $G$, a contradiction. Thus every caterpillar in $G[V \setminus S]$, and therefore every vertex $v \in V_1$, has at least one neighbor in $S$. If two vertices in $V_1$ have the same neighbor in $S$, then Rule 2 would apply to $G$, a contradiction. Thus every vertex in $V_1$ has a distinct neighbor in $S$, and so $|V_1| \leq |S| = k$. □

**Lemma 6.** *Let $(G, k)$ be a YES instance of the problem that is reduced with respect to Rule 3. For a set $X \subseteq V$, if $A \subseteq V \setminus X$ is such that every $v \in A$ has (i) at least two neighbors in $X$, and (ii) no neighbors outside $X$, then $|A| \leq \binom{|X|}{2}(k + 2)$.*

*Proof.* To prove this bound on $|A|$, we start by associating an integer $x_{ij} = 0$ with each pair of vertices $\{v_i, v_j\} \subseteq X$. We then go through the vertices of $A$, and for each vertex $u \in A$, we find a pair of vertices $\{v_i, v_j\} \subseteq N(u)$ such that $x_{ij} < (k+2)$, and increment this $x_{ij}$ by one. We will always be able to do this, or else Rule 3 would apply to vertex $u$, a contradiction. At the end of this process, $|A| = \sum_{\{v_i,v_j\}\subseteq X} x_{ij}$. But for each pair of vertices $\{v_i, v_j\} \subseteq X$, $x_{ij} \leq (k+2)$, and it follows that $|A| \leq \binom{|X|}{2}(k+2)$. $\square$

**Lemma 7.** *Let $(G = (V, E), k)$ be an instance of the problem that is reduced with respect to Rules 1 to 5, and let $S \subseteq V$ be such that $G[V \setminus S]$ has pathwidth at most one. Let $X \subseteq (V \setminus S)$ be the set of vertices in $(V \setminus S)$ that lie on the spines of nontrivial caterpillars in $G[V \setminus S]$. Then $|X| \leq 17k(k+2)$.*

*Proof.* Let $C_1, C_2, \ldots, C_p$ be the nontrivial caterpillars in $G[V \setminus S]$, and for $1 \leq i \leq p$, let $P_i = \langle v_1, v_2, \ldots, v_{r_i} \rangle$ be a path of the maximum length in $C_i$. It is sufficient to show that $\sum_{i=1}^{p} r_i \leq 17k(k+2)$. Let $\mathcal{C}_s = \{C_i \mid |P_i| \leq 8\}$ (the "small" caterpillars), and let $\mathcal{C}_l$ be the remaining, "large" caterpillars.

Each $C_i \in \mathcal{C}_s$ has at least one neighbor in $S$, or else Rule 1 would apply. Any one $v \in S$ can have neighbors in at most $k + 2$ different elements of $\mathcal{C}_s$, or else Rule 4 would apply to $v$ and its neighborhood. It follows that $|\mathcal{C}_s| \leq k(k+2)$, and so the total number of vertices that lie on the spines of the elements of $\mathcal{C}_s$ is at most $8k(k+2)$.

Now we consider the caterpillars in $\mathcal{C}_l$. Without loss of generality, let $\mathcal{C}_l = \{C_1, C_2, \ldots, C_{p'}\}$. For $1 \leq i \leq p'$, let $P'_i = \langle v_3, v_4, \ldots, v_{r_i-2} \rangle$. $P'_i$ can be thought of as containing *blocks* $B_1, B_2, \ldots$, where each block consists of five consecutive vertices in the path. More specifically, $B_1 = \langle v_3, v_4, \ldots, v_7 \rangle$, $B_2 = \langle v_8, v_9, \ldots, v_{12} \rangle$, and so on, until there are fewer than 5 vertices left. Consider any block $B_j$ on path $P'_i$. Let $C_j$ be the set of pendant vertices, not belonging to $P_i$, adjacent to the vertices of $B_j$ in $G[V \setminus S]$, and let $X_j = B_j \cup C_j$. Let $x, y$ be the two vertices belonging to $P_i \setminus B_j$ that are adjacent to the two end vertices of $B_j$ in $G[P_i]$, and let $x', y'$ be the two vertices belonging to $P_i \setminus B_j$ that are adjacent to $x, y$, respectively. Thus, for example, for $B_2$ defined as above we have $x = v_7, x' = v_6, y = v_{13}, y' = v_{14}$. Note that $x, y, x', y'$ as defined here are guaranteed to exist for each block $B_j$. Also note that for any $B_j$, $(X_j, x, y, x', y')$ as defined here satisfy the requirements of Rule 5 in $G[V \setminus S]$ with $X_j$ as $X$. So, if none of the vertices of $X_j$ is adjacent to any vertex of $S$, then $(X_j, x, y, x', y')$ would satisfy these requirements in $G$ as well, in which case Rule 5 would apply to $G$, a contradiction. It follows that in $G$, at least one vertex of $X_j$ has an edge to a vertex of $S$. By the same argument as above, there are at most $k(k+2)$ distinct blocks in $G[V \setminus S]$. It follows that the total number of vertices that lie on the spines of the elements of $\mathcal{C}_l$ is at most $9k(k+2)$.

Putting these together, the bound in the lemma follows. $\square$

**Lemma 8.** *Let $(G = (V, E), k)$ be a YES instance of the problem that is reduced with respect to Rules 1 to 5, and let $S \subseteq V; |S| \leq k$ be such that $G[V \setminus S]$ has pathwidth at most one. Let $P \subseteq (V \setminus S)$ be the set of pendant vertices in $G[V \setminus S]$ that have at least one neighbor in $S$. Then $|P| \leq 17(k+2)^2 k(2k-1)$.*

*Proof.* Let $T \subseteq (V \setminus S)$ be the set of vertices that lie on the spines of caterpillars in $G[V \setminus S]$. By Lemma 7, $|T| \leq 17k(k+2)$. Partition $T$ into $l$ parts $T = T_1 \uplus T_2 \uplus \cdots \uplus T_l$ where each $T_i; 1 \leq i < l$ contains exactly $k$ vertices, and $T_l$ contains the remaining at most $k$ vertices. Clearly $l \leq 17(k+2)$. For $1 \leq i \leq l$, let $P_i = \cup_{v \in T_i}(N(v) \cap P)$; then $P = \cup_i P_i$. For $1 \leq i \leq l$, setting $X = S \cup T_i, A = P_i$ and applying Lemma 6 we get $|P_i| \leq \binom{|S \cup T_i|}{2}(k+2) \leq \binom{2k}{2}(k+2) = k(2k-1)(k+2)$. Hence $|P| \leq l \cdot k(2k-1)(k+2) = 17(k+2)^2 k(2k-1)$. □